\documentclass[%
 reprint,
 amsmath,amssymb,
 aps,
]{revtex4-2}

\usepackage{graphicx}% Include figure files
\usepackage{dcolumn}% Align table columns on decimal point
\usepackage{bm}% bold math
\usepackage{float}
\usepackage{graphicx}
\usepackage{subcaption}
\usepackage{mathastext}
\usepackage{colortbl} % For row coloring
\usepackage{xcolor}   % Define light gray

\begin{document}

\preprint{APS/123-QED}

\title{Electron magnetization effects on carbonaceous dusty nanoparticles grown in $Ar/C_2H_2$ capacitively coupled nonthermal plasma}% Force line breaks with \\

\author{Bhavesh Ramkorun}
\email{bzr0051@auburn.edu}
\author{Saikat C. Thakur}%
\author{Edward Thomas, Jr.  }%
 
\affiliation{%
 Department of Physics, Auburn University, AL, USA
}%


\author{Ryan B. Comes}
\affiliation{Department of Materials Science and Engineering, University of Delaware, DE, USA}%
%\affiliation{ Third institution, the second for Charlie Author}%
%\affiliation{%Authors' institution and/or address\\This line break forced with \textbackslash\textbackslash}%


\date{\today}% It is always \today, today,

\begin{abstract}

Carbonaceous dusty nanoparticles spontaneously grow in nonthermal plasmas from a gas mixture of argon and acetylene. These particles levitate and grow within the bulk plasma for a duration known as the growth cycle ($T_c$), after which they gradually move away. In experiments operating at 500 milliTorr, the particles reach a maximum radius of approximately 250 nm for $T_c \sim$ 121 s. However, the introduction of weak magnetic fields reduces both the maximum radius and $T_c$. The modified electron Hall parameter ($H_e'$), which quantifies the degree of electron magnetization, increases linearly with the magnetic field strength, transitioning from unmagnetized electrons ($H_e' < 1$) to magnetized electrons ($H_e' > 1$). $T_c$ gradually decreases to around 40 s until $H_e' \sim 1$ at approximately 330 Gauss, after which it remains roughly constant for fields up to about 1020 Gauss. Additionally, with increasing magnetic field strength, the dust growth rate initially decreases to $H_e' \sim 1$, then increases slightly again. These results demonstrate that the onset of electron magnetization at can control the growth of nanoparticles from chemical precursors in nonthermal plasmas, which is relevant for industrial applications.

\end{abstract}

\maketitle

%\tableofcontents

\section{Introduction} \label{sec:introduction}

A dusty plasma is a type of plasma that, in addition to electrons, ions, and neutrals, also contains nanometer (nm) to micrometer-sized charged particles, known as dust \cite{shukla2001survey, beckers2023physics}. Although dust particles can be introduced to a plasma \cite{thomas1994plasma, jaiswal2019melting, chaubey2023controlling, chaubey2024controlling, kumar2024producing},
dust can also grow in non-thermal plasma ignited with argon (Ar) and reactive precursor gases such as acetylene ($C_2H_2$) \cite{mao2008new,stefanovic2017influence, groth2019spatio}, silane \cite{bouchoule1991particle, bouchoule1993particulate, qin2016laser}, aniline \cite{pattyn2018nanoparticle}, hexamethyldisiloxane \cite{garofano2015cyclic,garofano2019multi}, water vapor \cite{chai2015study, marshall2017identification} and titanium isopropoxide \cite{ramkorun2023growth, ramkorun2024introducing}. These studies have shown that the physical and chemical properties of dust can be controlled within the plasma. For example, the dust size distribution is generally monodisperse and increases linearly over time, allowing excellent control over the particle size \cite{groth2015kinetic, kovavcevic2009formation}. Furthermore, the chemical bonding properties can change with increasing radius \cite{kovacevic2012size}. Consequently, nanospheres can possibly be controlled and grown for various applications, such as quantum dots \cite{kortshagen2008plasma, boufendi2011dusty}.

Dust usually levitates in the plasma due to various forces, such as electric, gravitational, ion-drag, neutral-drag, and thermophoretic forces, which depend on the radii of the particles \cite{shukla2009colloquium, van2015fast}. Eventually, the dust grows to a maximum size within a growth cycle and acquires enough charge and mass to overcome these forces in the bulk plasma, causing them to move away and allowing a new generation of particle growth to begin. This process occurs cyclically as long as the plasma contains a mixture that includes reactive gases. Studies in the literature have measured the particle growth cycle from intensity variation in Rayleigh/Mie scattering \cite{kovavcevic2003infrared}, pressure, self-bias \cite{hundt2011real}, optical emission spectroscopy (OES) \cite{garofano2015cyclic}, metastable density \cite{sushkov2016metastable} and camera images of the dusty plasma \cite{jaiswal2020effect}. Generally, cyclic growth is observed in various capacitively coupled dishcarges with several of the aforementionned reactive gases. 

Recent studies have shown that cycle time decreases when a weak magnetic field of up to 1000 Gauss is present. For example, Couedel et al. grew carbonaceous dust in Ar/$C_2H_2$ discharges and measured a cycle time of 40-60 seconds without any magnetic fields, which was reduced to 15-20 seconds during fields of 0.032-0.1 T (320-1000 Gauss) \cite{couedel2019influence}. The magnetic field did not seem to affect the material properties, as detected by Raman spectroscopy, with the samples consistently displaying a noncrystalline graphitic carbonaceous nature. Similarly, Ramkorun et al. observed a decrease in cycle time for carbonaceous and titanium dioxide (titania) dusty plasma \cite{ramkorun2024comparing}. The cycle time for carbonaceous and titania decreased from 115 ± 5 seconds and 77 ± 4 seconds (without magnetic field) to 39 ± 1 second and 32 ± 3 seconds (at 500 Gauss), respectively. From Langmuir probe measurements of the background plasma's floating potential, they suggested that changes in the radial behavior of the electric field might be responsible for the changes in cycle time. At magnetic fields greater than 0.5 T (5000 Gauss), the formation of filamentation in magnetized plasma appears to disrupt the particle growth cycle such that no cycle time is observed \cite{jaiswal2020effect, couedel2019influence}.

There has yet to be a study investigating how the cycle time depends on varying magnetic field strengths. The objective of this study is to grow dust in an $Ar/C_2H_2$ capacitively coupled plasma discharge and to conduct a comprehensive experimental investigation of particle growth behavior both in the absence and presence of a ``weak" magnetic field, ranging from approximately 18 to 1020 Gauss. We measure the cycle time through the intensity variation in the OES of the background Ar plasma. Our findings indicate that the cycle time decreases rapidly with increasing magnetic field strength until the electrons become fully magnetized, after which the cycle time stabilizes at a minimum value. Additionally, we observe that the growth rate of nanoparticles during the initial cycle varies with the magnetic field strength.

\section{Experimental methods}

\begin{figure*}[b]
    \centering
    \includegraphics[width=\textwidth]{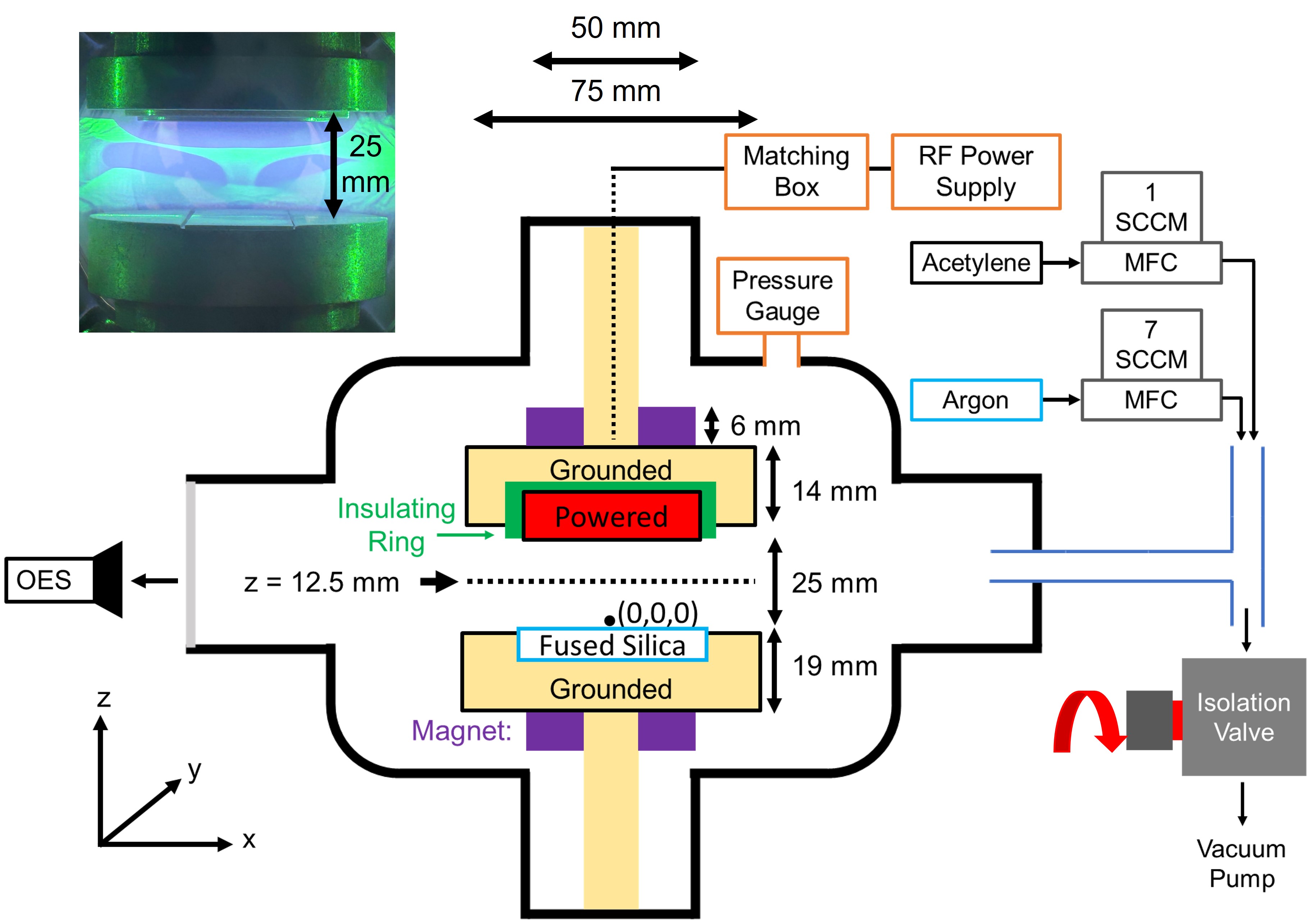}
    \caption{
        Schematic of the experimental setup (not to scale). Parallel electrodes are positioned with a pair of magnet rings symmetrically placed at $z = 12.5$ mm. The system is housed inside a 6-way cross vacuum chamber, where gases flow to sustain the plasma, which is ignited by an RF power source. The inset (top left) shows a camera image of a dusty plasma at 460 Gauss, with the dust illuminated by a green laser sheet.}
    \label{fig:exptSetUp}
\end{figure*}

The particle growth chamber was a standard stainless steel 6-way cross with ISO 100 flanges. The top and bottom ports were used to support the electrodes to ignite a capacitively coupled plasma. The electrodes each had a diameter of 75 mm. The top electrode had a 14 mm thickness, while the bottom electrode had a 19 mm thickness. Their separation (electrode spacing) was 25 mm. A part of the top electrode (50 mm) was powered by an RF generator (13.56 MHz), while the rest acted as a grounded counter electrode, separated by an insulating alumina ring. The bottom electrode was grounded to confine the plasma between the electrodes. It also had a slot to fit a fused silica slide to collect dust after growth for subsequent SEM imaging. The configuration of the electrodes and the location of the magnets inside the chamber are schematically shown in Fig. \ref{fig:exptSetUp}.

Ar was used to ignite the background plasma, and $C_2H_2$ served as the reactive precursor gas for growing carbonaceous dusty nanoparticles. Two mass flow controllers (MFC) were used to individually regulate the flow of Ar and $C_2H_2$ at 7 and 1 standard cubic centimeters per minute (SCCM), respectively. Initially, the base pressure of the chamber was 2 milliTorr (mTorr). After the gases began flowing, an isolation valve connecting the chamber to the vacuum pump was adjusted to reduce the conductance of gases out of the chamber, thus raising the chamber pressure to 500 mTorr. The power supply was then ignited at 60 Watts (W), with the reflected power kept at a minimal 1 W. This formed the carbonaceous dusty plasma, an example picture of which, at 460 Gauss, is also shown in Fig. \ref{fig:exptSetUp}.

A broadband spectrometer (Avantes AvaSpec-ULS4096CL-EVO), with resolution of 0.59 nm, was used to capture the optical emission spectrum of Ar I at 763.5 nm over several minutes to determine the cycle time $\left(T_c\right)$ from its intensity variation. This corresponds to neutral Ar with the electronic transition responsible for light emission occurring between the metastable states of $3s^2 3p^5\left( {}^2P^\circ_{3/2} \right)4s$ and $3s^2 3p^5\left( {}^2P^\circ_{3/2} \right)4p$
 \cite{NIST_ASD, PhysRevA.39.2461, norlen1973wavelengths}. This line was chosen because it was relatively strong, isolated, and had no other peaks blending. It has been commonly used to monitor the background of dusty plasma \cite{samsonov1999instabilities, couedel2013growth, bouchoule1994high}. There was no line ratio experiment;  the cyclic variation in OES intensity was used simply to extract $T_c$. The particles were then grown and collected at four different time intervals within the first cycle at $T_c/4$, $T_c/2$, $3T_c/4$, and $T_c$.

\begin{figure*}[htbp]
    \centering
    % First row of subfigures
    \begin{subfigure}{0.49\textwidth}
        \centering
        \includegraphics[width=\textwidth]{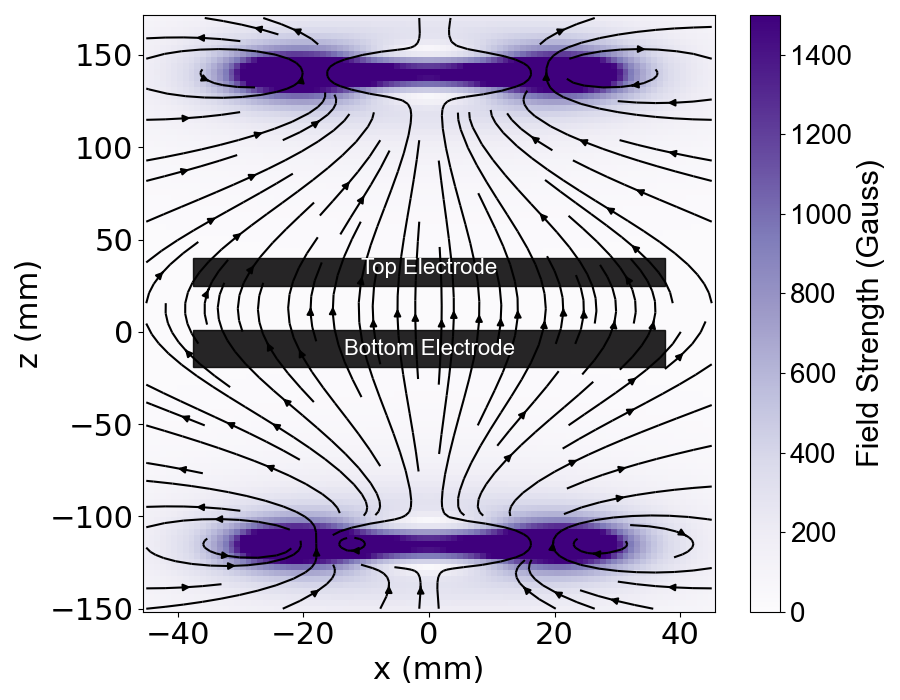}
        \caption{}
        \label{fig:a}
    \end{subfigure}
    \hfill
    \begin{subfigure}{0.49\textwidth}
        \centering
        \includegraphics[width=\textwidth]{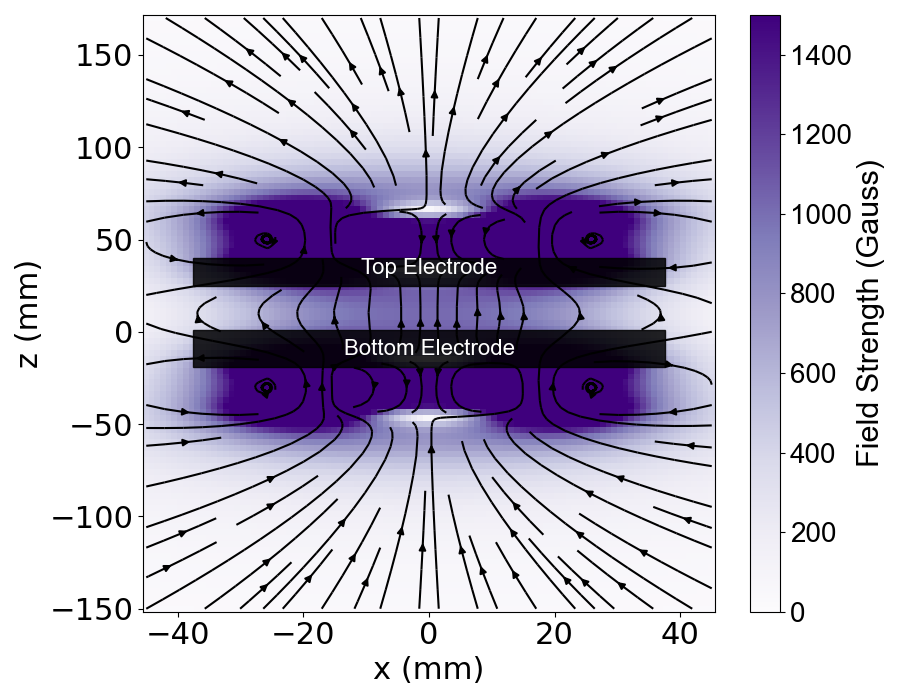}
        \caption{}
        \label{fig:e}
    \end{subfigure}

    \vspace{0.25cm} % Adjust spacing between rows

    % Second row of subfigures
    \begin{subfigure}{0.49\textwidth}
        \centering
        \includegraphics[width=\textwidth]{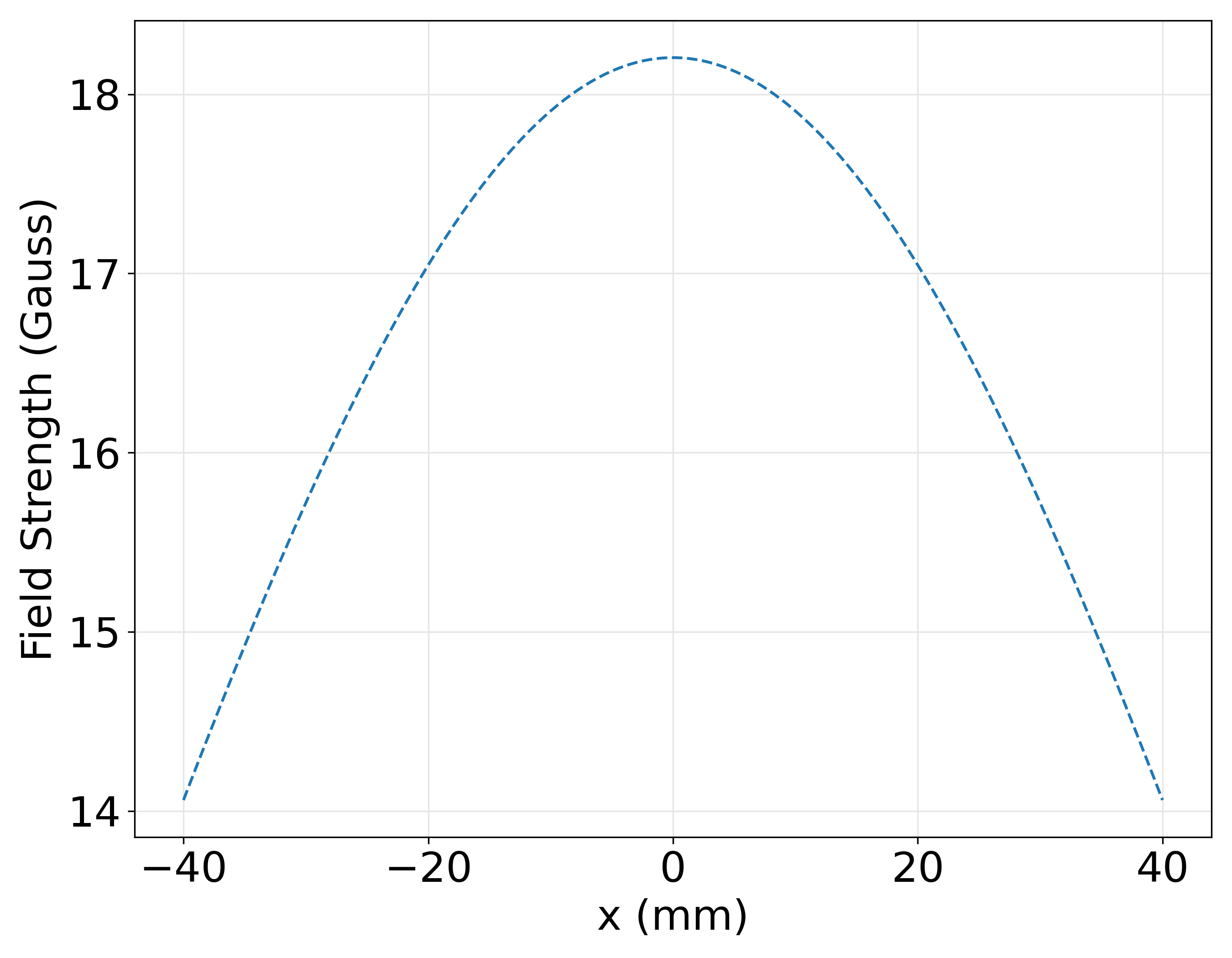}
        \caption{}
        \label{fig:c}
    \end{subfigure}
    \hfill
    \begin{subfigure}{0.49\textwidth}
        \centering
        \includegraphics[width=\textwidth]{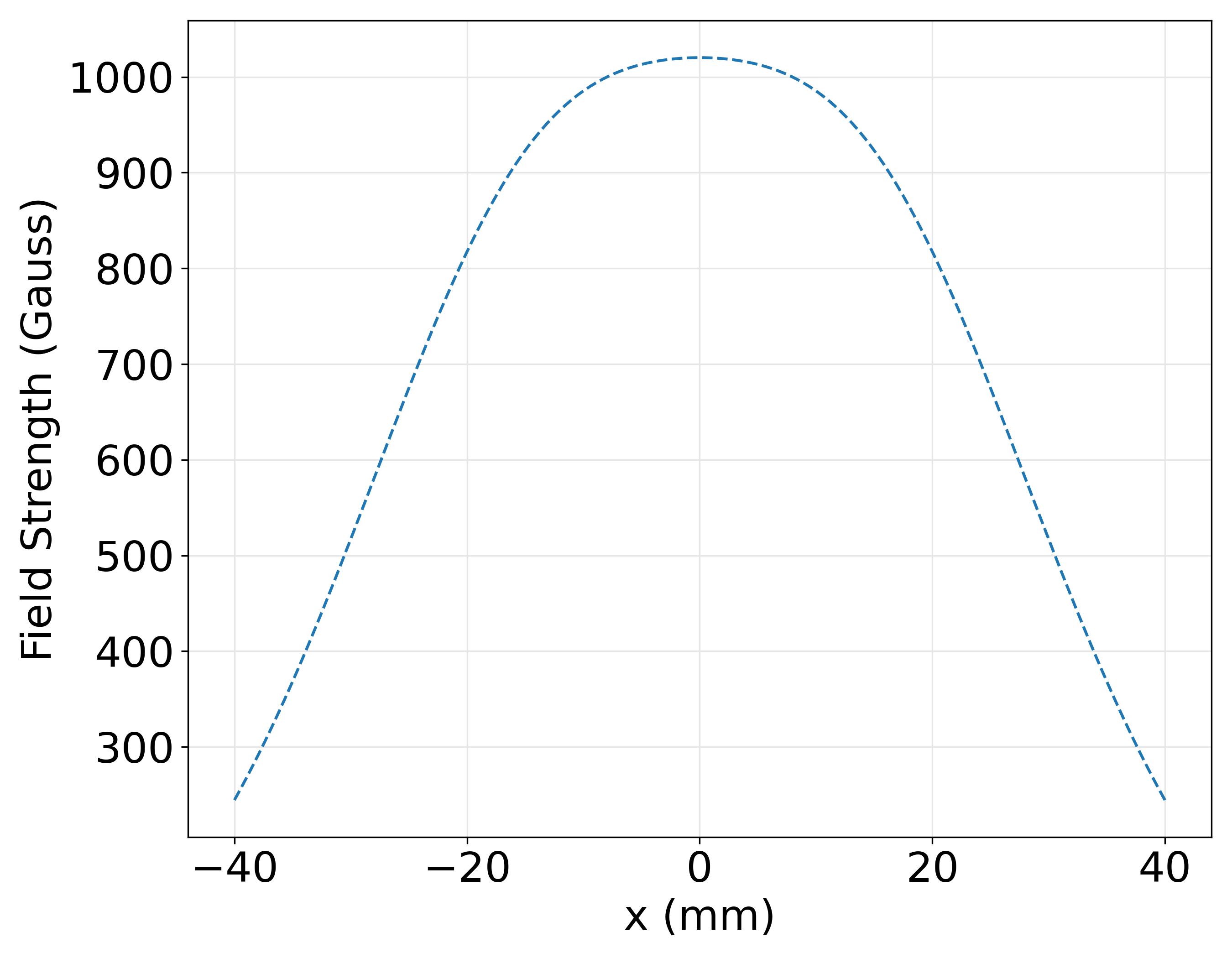}
        \caption{}
        \label{fig:d}
    \end{subfigure}

    \caption{Example of the 2-D magnetic field between the electrodes. (a) At the largest magnet separation (255 mm), the field reaches a maximum of approximately 18 Gauss at the center of the electrodes ($x = 0$ mm, $z = 12.5$ mm). (b) At the closest separation (66 mm), two additional pairs of magnets are stacked, producing a maximum field of approximately 1020 Gauss at the same location. (c) and (d) show the magnetic field strength along $x$ at $z = 12.5$ mm for 18 Gauss and 1020 Gauss, respectively.}
    \label{fig:field_strength}
\end{figure*}

To produce magnetic fields in the experiments, pairs of permanent magnet rings made of neodymium iron boron (NdFeB) of grade N42 were used. These magnets were placed symmetrically from the center of the electrode spacing (at $z = 12.5$ mm) to maximize the magnetic field strength at the center of the plasma, where the $T_c$ would be measured. The maximum separation between a pair of magnets (255 mm) created a maximum field strength of approximately 18 Gauss at the geometric center of the electrodes. The spacing between the magnets was gradually decreased in each experiment to produce various field strengths. The minimum separation between a pair of magnets (64 mm) created a maximum field strength of approximately 460 Gauss. At this separation, either one or two additional magnets were stacked on the original pair to increase the magnetic field strength to either 790 or 1020 Gauss, respectively. The 2-D magnetic field strength was calculated using Magpylib \cite{ortner2020, pizzey2021}. Two examples of magnetic field strength in 2-D plots are shown in Fig. \ref{fig:field_strength} for 18 and 1020 Gauss, together with a line plot of magnetic field strength along the middle at $z = 12.5$ mm for a range of x spanning the diameter of the electrodes. A summary of the different magnetic field strengths, the number of pairs of magnets, and their locations at the top and bottom is provided in Table \ref{tab:magnetic_field}.

\begin{table}[h]
    \caption{Magnetic field strengths and corresponding locations of magnet pairs. Field strengths highlighted in light gray were used in the particle growth experiment to determine the growth rate. 0.5 Gauss represents Earth's background field with no external magnets.}
    \label{tab:magnetic_field}
    \centering
    \begin{ruledtabular}
    \begin{tabular}{c c c c}
        \textbf{Field Strength} & \textbf{Pairs of} & \textbf{Bottom} & \textbf{Top} \\
        \textbf{(Gauss)} & \textbf{Magnets} & \textbf{z (mm)} & \textbf{z (mm)} \\
        \hline
        \cellcolor{lightgray} 0.5  & 0 & -    & -   \\
        &  & & \\
        \cellcolor{lightgray} 18   & 1 & -115 & 140 \\
        &  & & \\
        26   & 1 & -100 & 125 \\
        &  & & \\
        \cellcolor{lightgray} 73   & 1 & -65  & 90  \\
        &  & & \\
        87   & 1 & -60  & 85  \\
        &  & & \\
        \cellcolor{lightgray} 120  & 1 & -52  & 77  \\
        &  & & \\
        195  & 1 & -39  & 64  \\
        &  & & \\
        \cellcolor{lightgray} 233  & 1 & -35  & 60  \\
        &  & & \\
        335  & 2 & -39  & 64  \\
             &   & -47  & 72  \\
             &  & & \\
        \cellcolor{lightgray} 460  & 1 & -22  & 42  \\
        &  & & \\
        \cellcolor{lightgray} 790  & 2 & -22  & 42  \\
             &   & -30  & 50  \\
             &  & & \\
        \cellcolor{lightgray} 1020 & 3 & -22  & 42  \\
             &   & -30  & 50  \\
             &   & -38  & 68  \\
    \end{tabular}
    \end{ruledtabular}
\end{table}

\section{Results} \label{sec:experimental}
\subsection{Cycle time}
\begin{figure}[t]
    \centering
    \begin{subfigure}[t]{0.9\linewidth}
        \centering
        \includegraphics[width=\linewidth]{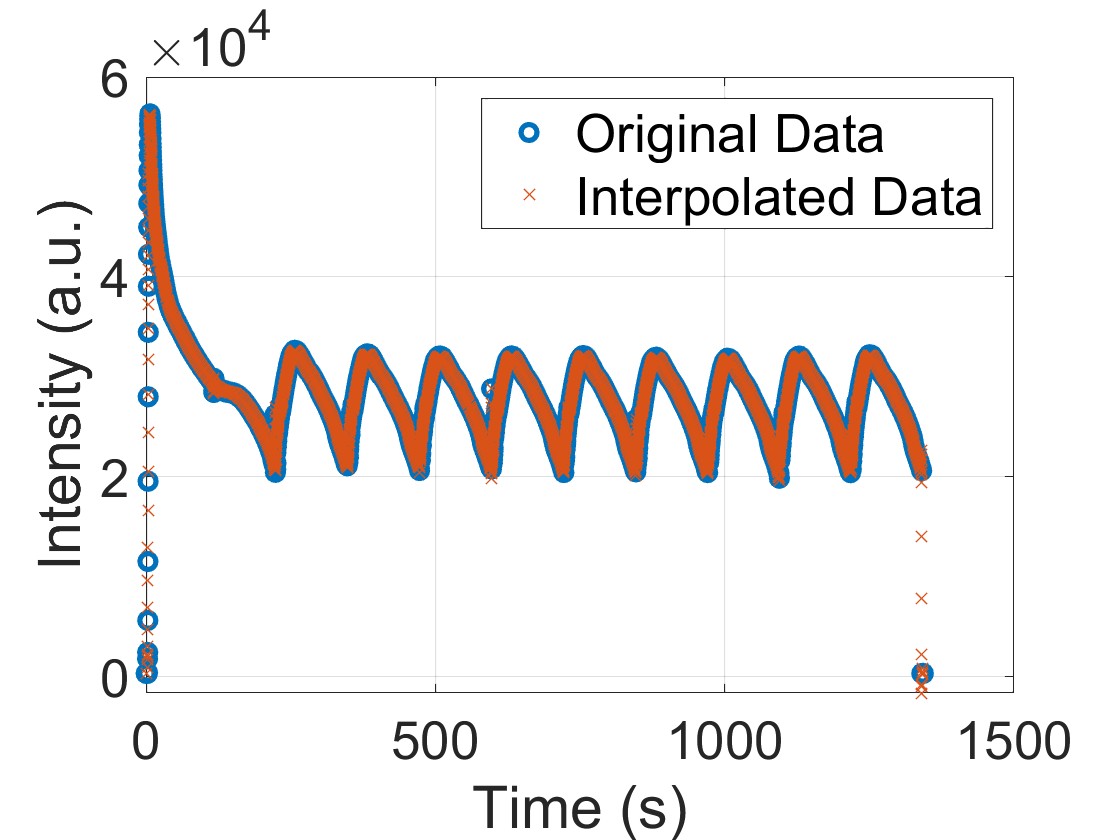}
        \caption{}
        \label{fig:oes_cycle_interpolated}
    \end{subfigure}

    \vspace{0.3cm}

    \begin{subfigure}[t]{0.9\linewidth}
        \centering
        \includegraphics[width=\linewidth]{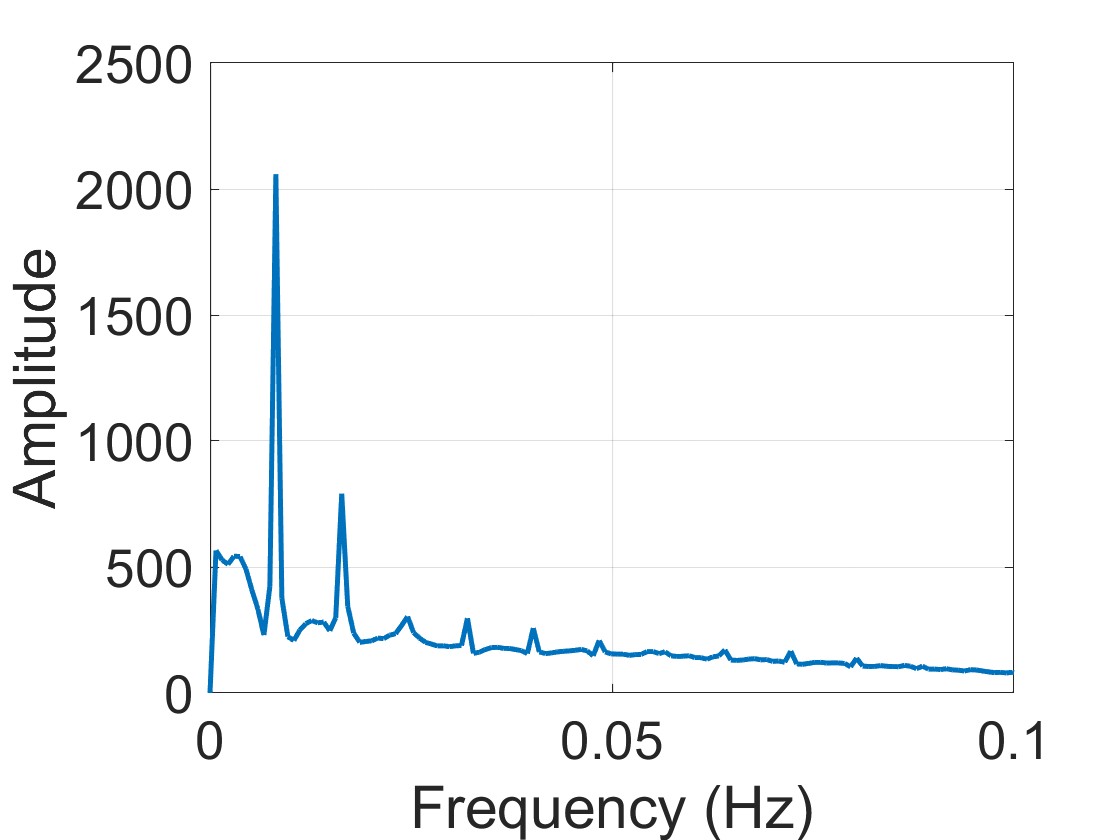}
        \caption{}
        \label{fig:oes_fft}
    \end{subfigure}

    \vspace{0.3cm}

    \begin{subfigure}[t]{0.9\linewidth}
        \centering
        \includegraphics[width=\linewidth]{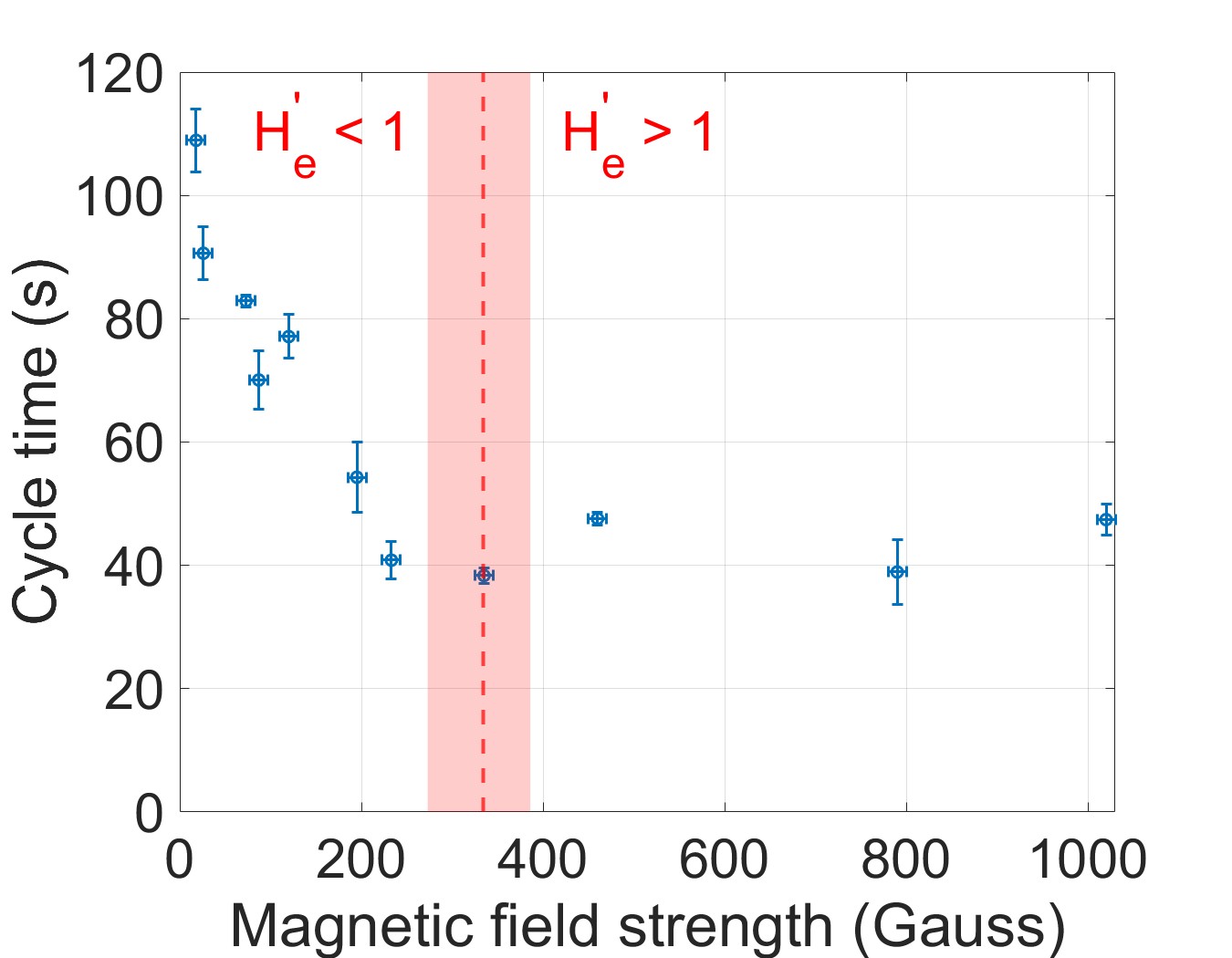}
        \caption{}
        \label{fig:cycle_time_field}
    \end{subfigure}

    \caption{Extracting cycle time from OES of the dusty plasma: (a) Temporal variation of Ar I OES intensity at 763.5 nm with spline interpolation. (b) Frequency spectrum obtained from FFT analysis of (a) after subtracting the average from all the data points. (c) Cycle time (inverse of frequency with maximum intensity) versus magnetic field strength, showing a decrease until 330 Gauss, after which it plateaus.}
    \label{fig:overall}
\end{figure}

Previous studies have demonstrated that the cyclic variation of the OES intensity mirrors the cyclic growth cycle of dusty plasma \cite{ramkorun2024introducing, garofano2015cyclic, ramkorun2024comparing}. Therefore, the cyclic variation of Ar I at 763.5 nm was measured for at least 8 cycles to determine the cycle time of the nano dusty plasma. An example of the cyclic variation of the OES intensity at 18 Gauss is shown in Figure \ref{fig:oes_cycle_interpolated}. The data were collected at time intervals of approximately 200 milliseconds and a spline interpolation was used to interpolate the data at every 100 milliseconds. From the interpolated data, a fast Fourier transform (FFT) was performed to determine the peak frequency of the cycles, as shown in Figure \ref{fig:oes_fft}. The average intensity  was subtracted from each data point in order to extract any noise from the DC level.

At least three sets of data were collected and averaged for each magnetic field configuration to determine the cycle time as a function of magnetic field strength. The cycle time was calculated as the inverse of the peak frequency with maximum intensity. The results are shown in Figure \ref{fig:cycle_time_field}. Initially, without any external magnetic field (0.5 Gauss from Earth's magnetic field), $T_c$ was 121 $\pm$ 7 s. This gradually decreased to 40 $\pm$ 3 s at 233 Gauss and continued to fluctuate between 38 and 47 s as the magnetic field strength increased from 335 to 1020 Gauss. Overall, we observe that $T_c$ gradually decreased between 0.5 and 233 Gauss and remained relatively constant up to 1020 Gauss.

\subsection{Growth Rate}

\begin{figure}[t]
    \centering
    % First Image
    \begin{subfigure}[b]{0.9\linewidth}
        \centering
        \includegraphics[width=\linewidth]{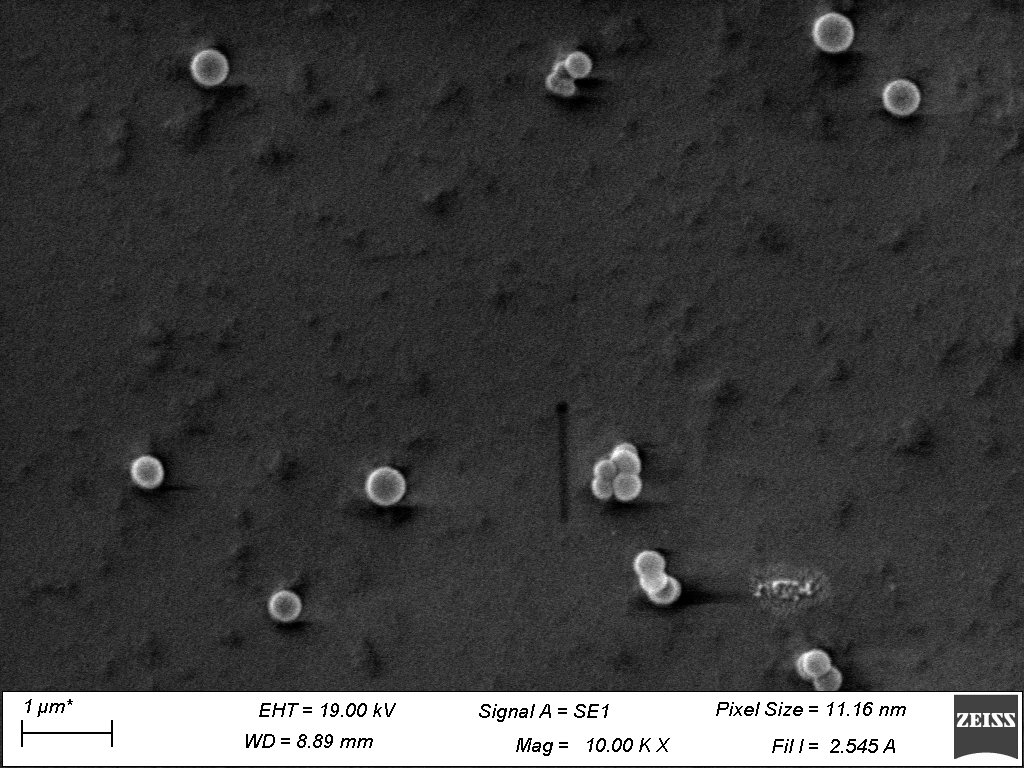}
        \caption{}
        \label{fig:SEM_a}
    \end{subfigure}

    \vspace{0.3cm}

    % Second Image
    \begin{subfigure}[b]{0.9\linewidth}
        \centering
        \includegraphics[width=\linewidth]{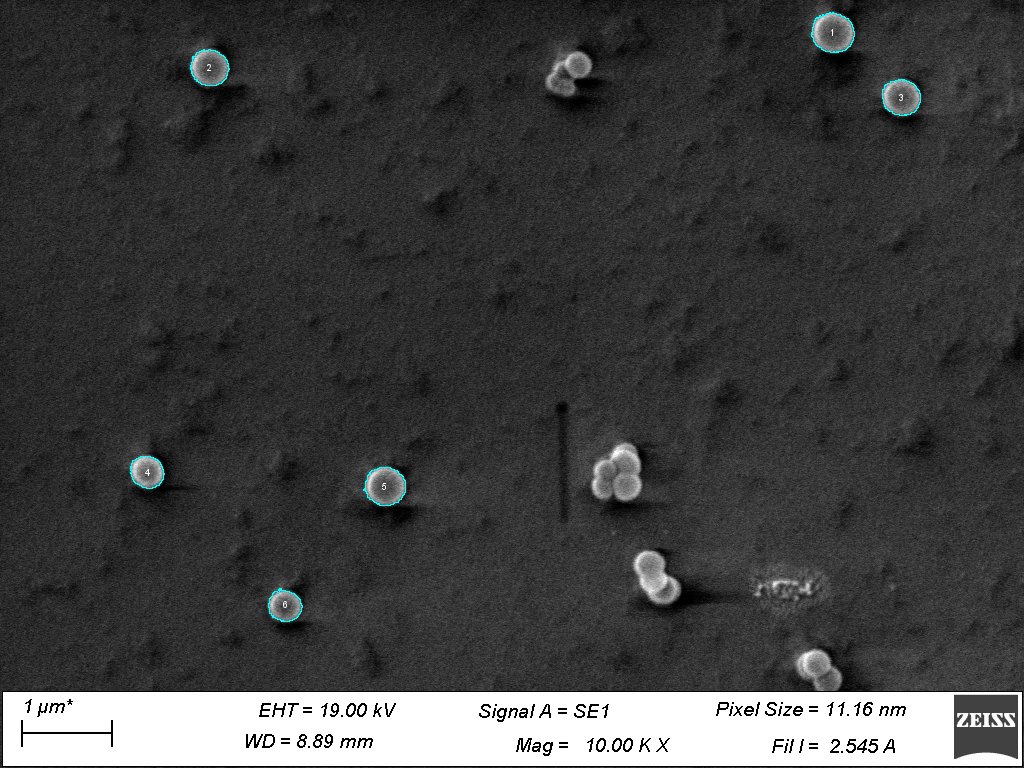}
        \caption{}
        \label{fig:SEM_b}
    \end{subfigure}

    \caption{Scanning electron microscope (SEM) images of nanoparticles grown for $T_c = 47$ s at 1020 Gauss: (a) Original SEM image. (b) The same image with isolated particles highlighted for size distribution analysis.}
    \label{fig:sem_images}
\end{figure}

Once $T_c$ was established, dust was grown again and collected at four intervals within the first cycle: $T_c/4$, $T_c/2$, $3T_c/4$, and $T_c$. The goal was to collect the dust on the fused silica slide at the bottom electrode, image it using a scanning electron microscope (SEM), and determine the particle size distribution to assess the growth rate within the first cycle. Only clear and isolated particles were selected for size extraction. For example, Figure \ref{fig:sem_images} shows nanoparticles collected at 1020 Gauss after growing for $T_c \sim 47$ s. In Figure \ref{fig:SEM_a}, the original SEM image displays several isolated particles as well as particles sticking together. However, in Figure \ref{fig:SEM_b}, only isolated nanoparticles, circled in teal, were chosen to determine the particle size distribution. This process was repeated for all imaged nanoparticles. When levitating in the plasma, nanoparticles are negatively charged and repel each other. However, once the plasma is turned off, they can experience charge fluctuations within milliseconds \cite{chaubey2023controlling, chaubey2024controlling}. This might cause them to stick together while falling onto the fused silica slide for collection. These particles were excluded from the size distribution analysis because their shapes and edges could not be clearly delineated to determine radii.

\begin{figure}[t]
    \centering
    \begin{subfigure}[b]{0.9\linewidth}
        \centering
        \includegraphics[width=\linewidth]{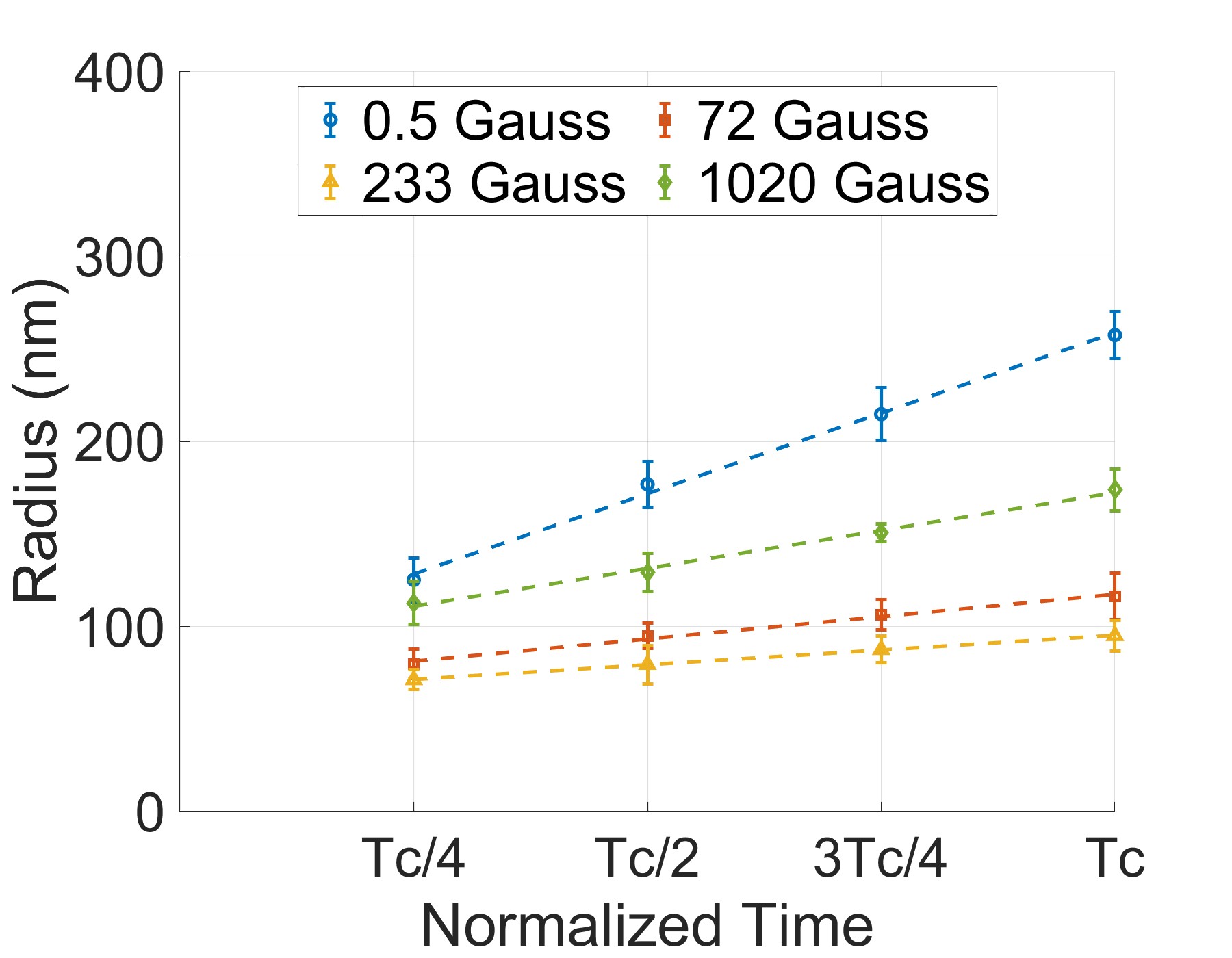}
        \caption{}
        \label{fig:growth_rate_per_cycle}
    \end{subfigure}

    \vspace{0.3cm}

    \begin{subfigure}[b]{0.9\linewidth}
        \centering
        \includegraphics[width=\linewidth]{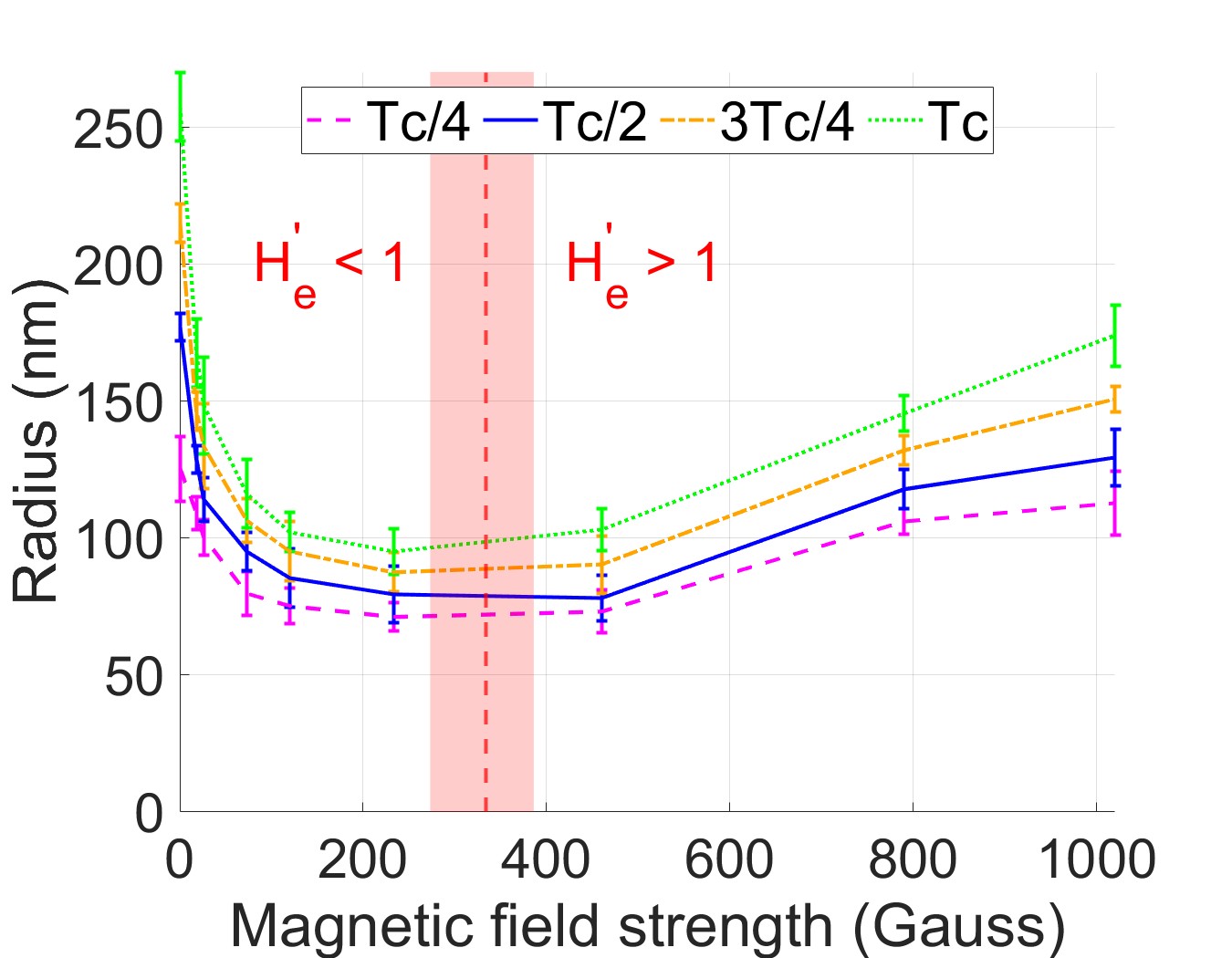}
        \caption{}
        \label{fig:max_radius_Tc}
    \end{subfigure}

    \vspace{0.3cm}

    \begin{subfigure}[b]{0.9\linewidth}
        \centering
        \includegraphics[width=\linewidth]{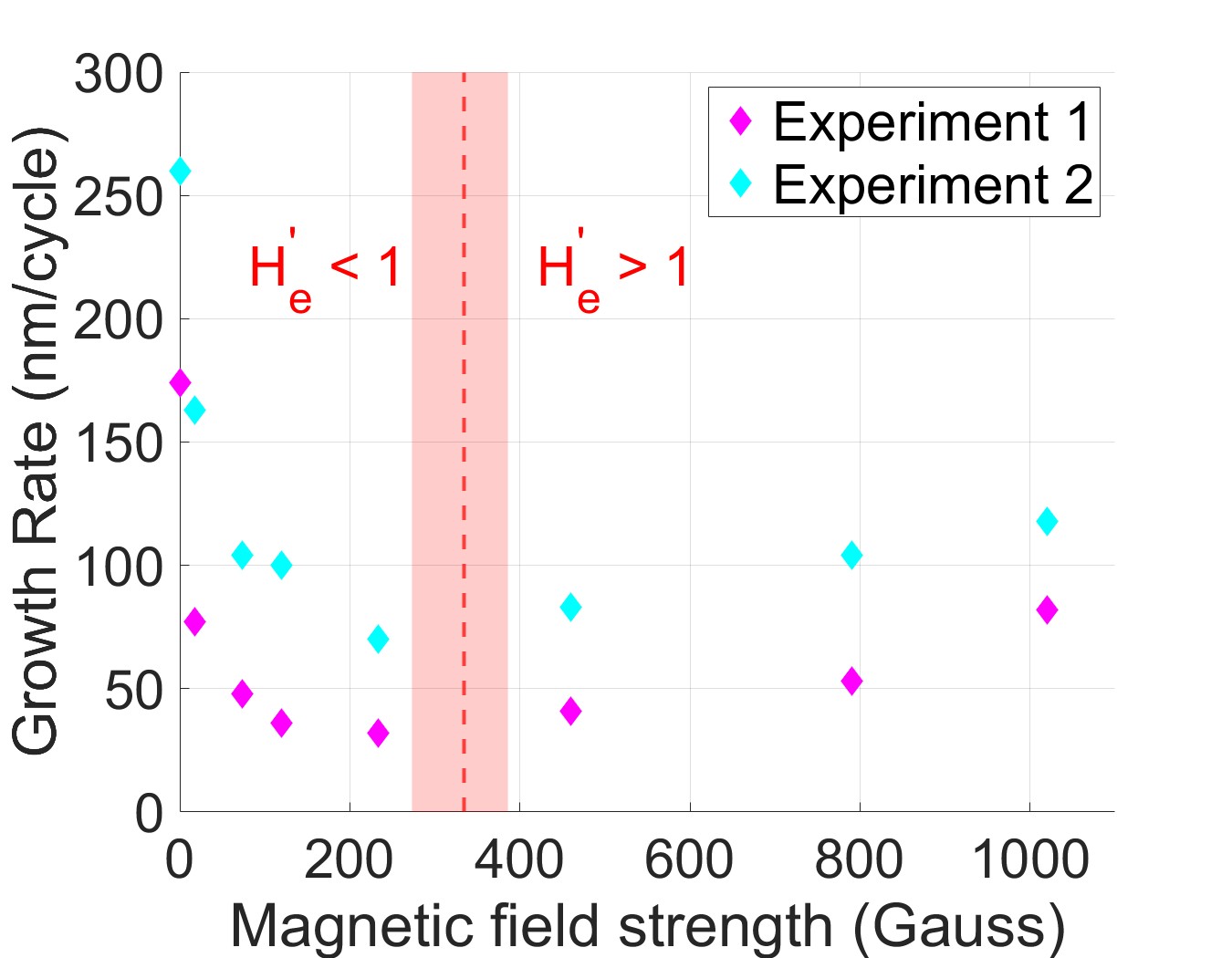}
        \caption{}
        \label{fig:growth_rate_third}
    \end{subfigure}

    \caption{Particle size distribution from SEM at four time intervals within one growth cycle at varying magnetic fields (Gauss). 
    (a) Changes in particles' radii from the first cycle. 
    (b) Radius as a function of magnetic field for each time interval. 
    (c) Growth rate within the first cycle (slope of (a)) as a function of magnetic field strength.}
    \label{fig:growth_rate_overall}
\end{figure}

The general results for the particle growth rate are shown in Fig. \ref{fig:growth_rate_overall}. We analyze the growth rate per cycle, meaning the growth rate of particles within a single cycle for each magnetic field configuration. We determined the growth rate within only the first cycle, because particles can exhibit a bimodal size distribution in subsequent cycles \cite{ramkorun2024introducing, ramkorun2024comparing}.  Examples of the results are shown in Fig. \ref{fig:growth_rate_per_cycle}. In this figure, we observe that the maximum particle radius ($\sim$ 250 nm) and growth rate ($\sim$ 178 nm/cycle), as determined by the slope of the graph, were achieved for particles grown without an external magnetic field.

In Fig. \ref{fig:max_radius_Tc}, the particle radius is shown as a function of magnetic field strength for each section of the cycle time. The size decreases with increasing field strength up to about 233 Gauss, after which it slightly increases with further increases in magnetic field strength. In Fig. \ref{fig:growth_rate_third}, the growth rate, extracted from Fig. \ref{fig:growth_rate_per_cycle}, is plotted as a function of magnetic field strength. We observe that the growth rate initially decreases to approximately 30 nm/cycle at 233 Gauss but gradually increases again to approximately 82 nm/cycle at 1020 Gauss. These results indicate that the particle size and growth rate in nonthermal plasmas initially decrease in the presence of a magnetic field and can be further controlled by varying the field strength. The data in Fig. \ref{fig:growth_rate_per_cycle} and Fig. \ref{fig:max_radius_Tc} are from one experiment, which was repeated twice in Fig. \ref{fig:growth_rate_third}, displaying similar growth patterns.

\section{Discussion}

To interpret the measurements presented in Sec. III, it is necessary to summarize how the particle growth is altered.  We start with an examination of  the maximum size of the grown particles in a single cycle.  The experimental results show: (a) that the maximum size obtained by the particles decreases with increasing magnetic field, and (b) there is a rapid reduction in the growth rate (in nm/cycle) with increasing magnetic field, up to B $\sim$ 250 Gauss.  This combination of factors provides some initial evidence that the presence of the magnetic field has  altered both the force balance on the particle that confines them in the plasma and the reaction rates that lead to particle formation.

In terms of the force balance, it is known that the zero order force balance that levitates the particles in the plasma is between determined by the electric and gravitational forces on the dust particles \cite{shukla2009colloquium, van2015fast}.  The important question is how might the presence of the magnetic field alter the distribution of the electric potential and, subsequently, the electric fields in the plasma. It is possible that the spatial distribution of electrons changed with the increasing magnetic field, thereby altering the plasma ambipolar electric field.

Let us consider why the behavior of electrons might change in the presence of magnetic fields. The modified Hall parameter, $\left(H'_{e}\right)$ given by Eq. \ref{eq:hall_param}, which is the ratio of the electron gyrofrequency ($\omega_{e}$) to the electron-neutral collision frequency ($\nu_{n,e}$), quantifies electron magnetization in a plasma exposed to magnetic fields \cite{thomas2015magnetized, hall2018methods, williams2022experimental}. The thermal velocity of the electron $\left(v_{\text{th}, e}\right)$ is given by Eq. \ref{eq:therm_v_elec}. When $H'_{e} > 1$, the electrons are magnetized, i.e. they are more likely to complete a full gyro-orbit around a magnetic field line before colliding with neutral species. It is possible that this changes the spatial distribution of electrons in the plasma, and thus altering the electric profile of the background plasma.

\begin{equation}
    H'_{e} = \frac{\omega_{e}}{2\pi \nu_{n,e}} =
\frac{\frac{q_{e} B}{m_{e}}}{N_n \sigma v_{\text{th}, e}}
\label{eq:hall_param}
\end{equation}

\begin{equation}
    v_{\text{th}, e} = \sqrt{\frac{8 k T_{e}}{\pi m_{e}}}
    \label{eq:therm_v_elec}
\end{equation}

The red line in Fig. \ref{fig:cycle_time_field} represents $B = 334$ Gauss, where $H'{e} \approx 1$, assuming an electron temperature of 3 eV, based on our previous measurements for both magnetized and unmagnetized plasma \cite{ramkorun2024comparing}. However, variations in electron temperature between 2 and 4 eV would shift the magnetic field range for $H'_{e} = 1$ from 273 to 386 Gauss, hence the red-colored band range of the magnetic field. To the right of the band ($H'_{e} > 1$), electron dynamics are expected to be dominated by the presence of a magnetic field. To the left of the band ($H'_{e} < 1$), electron dynamics are expected to be dominated by collisions. With increasing magnetic field strength beyond $H'_{e} > 1$, the spatial distribution of electrons in the plasma will be governed by the magnetic field, while ions will remain unmagnetized. We observe that $T_c$ and particle radii gradually decrease until $H'_{e} \approx 1$. 

\textbf{}

Three changes might be happening in the dusty plasma that could explain the observed phenomena. First, as electrons transition with increasing magnetic field strength, the electric potential of the plasma is modified. This alters the electric field and the force required for particle confinement, potentially explaining the reduced particle radii observed in the presence of magnetic fields. Second, the new spatial distribution of electrons might affect the charging of the dust particles. The particle charge is crucial for the force balance that levitates them in the plasma, and any changes in charging could be causing the reduced particle size and decreased cycle time ($T_c$). Third, increasing magnetic field strengths might affect the electron density in the dusty plasma. Dust acts as a sink for electrons, reducing both electron density and metastable argon density \cite{sushkov2016metastable}. This reduction may impact particle coagulation and agglomeration during growth, thus explaining changes in the particle growth rate. Decoupling these three factors—changes in the electric field, dust charging, and electron density—is beyond the scope of this work. However, understanding their behavior is crucial and encouraged for further studies on particle growth in magnetized plasmas.

\section{Conclusion}

In conclusion, we successfully grew carbonaceous dusty nanoparticles in a capacitively coupled nonthermal plasma under the influence of weak magnetic fields ranging from 18 to 1020 Gauss. Particles grown without additional external magnetic fields (0.5 Gauss) exhibited the largest radii, longest cycle times, and fastest growth rates. These parameters gradually decreased until the electrons became magnetized $\left(H'{e} = 1\right)$. Beyond this point, up to 1020 Gauss, the cycle time plateaued and the growth rate and particle size slightly increased, though not to the same extent as at 0.5 Gauss. These findings demonstrate that weak magnetic fields can change the behavior of electrons in nonthermal plasma which may impact material synthesis from gaseous chemical precursors. 

%%%%%%%%%%%%%%%%%%%%%%%%%%%%%%%%%%%%%%%%%%%
 \section*{Acknowledgement}
This work was supported with funding from the NSF EPSCoR program (OIA-2148653), and the U.S. Department of Energy – Plasma Science Facility (SC-0019176).  The authors express gratitude for the technical assistance provided by Mr. Cameron Royer, Dr. Mohtadin Hashemi, Dr. Md Fahim Salek, and Mr. Jeffrey Estep in facilitating the implementation of this research project.

%%%%%%%%%%%%%%%%%%%%%%%%%%%%%%%%%%%%%%%%%%%
\section*{Declaration of competing interest}

None.

%%%%%%%%%%%%%%%%%%%%%%%%%%%%%%%%%%%%%%%%%%%
\section*{Data Availability}
The data that support the findings of this study are available
from the corresponding author upon reasonable request.

%\nocite{*}



\end{document}